\documentclass[aps,pra,twocolumn,showpacs,reprint,groupedaddress,superscriptaddress,longbibliography]{revtex4-1}

\usepackage[english]{babel}
\usepackage{float}
\usepackage[caption=false]{subfig}
\usepackage{amsfonts}
\usepackage{amssymb}
\usepackage{mathbrack}
\usepackage{siunitx}
\usepackage{mathtools}
\usepackage{bbold}
\usepackage{physics}
\usepackage{mwe}
\usepackage{amsthm,enumitem}
\usepackage{tikz}
\usepackage{ulem}
\usetikzlibrary{arrows,arrows.meta,shapes}
\usetikzlibrary{positioning}
\pdfoutput=1
\usepackage{dcolumn}
\usepackage{bm}
\usepackage[utf8]{inputenc}
\usepackage{graphics,graphicx}
\usepackage{parskip}
\usepackage{amsmath,amssymb}
\usepackage{multirow}
\usepackage{fancyhdr}
\usepackage{xcolor}
\usepackage{placeins}
\usepackage[title]{appendix}
\usepackage{wasysym}
\usepackage{url}
\usepackage{braket}
\usepackage{siunitx}
\usepackage{upgreek}
\usepackage[caption=false]{subfig}

\begin{document}

\title{Computational Study of the Spectral Behaviour of Different Isospectrally Patterned Lattices}

\author{Peter Schmelcher}
\email{peter.schmelcher@uni-hamburg.de}
\affiliation{Zentrum f\"ur Optische Quantentechnologien, Fachbereich Physik, Universit\"at Hamburg, Luruper Chaussee 149, 22761 Hamburg, Germany}
\affiliation{The Hamburg Centre for Ultrafast Imaging, Universit\"at Hamburg, Luruper Chaussee 149, 22761 Hamburg, Germany}

\date{\today}

\begin{abstract}
We perform a computational spectral analysis of different isospectrally patterned lattices (IPL). Having been
introduced very recently, the lattice Hamiltonian of IPL consist of coupled cells which possess all the same set of
eigenvalues. The latter is achieved in a controllable manner by parametrizing the cells via the
phases of the involved orthogonal (or unitary) transformations. This opens the doorway of systematically designing
lattice Hamiltonians with unique properties by choosing correspondingly varying phases across the lattice.
Here we focus on two-dimensional cells and explore symmetric as well as asymmetric IPL w.r.t. a given center.
A tunable fraction of localized vs. delocalized eigenstates belonging to the three subdomains of the corresponding
energy bands is demonstrated and analyzed with different measures of localization. In the asymmetric case the
center of localization can be shifted arbitrarily by shifting the underlying phase grid. Introducing a complete
phase revolution leads for low and high energies to two well-separated branches of localized states which finally
merge with increasing energy into the branch of delocalized states. Remarkably, the localized states appear in
near-degenerate pairs and this near-degeneracy is lifted upon entering the delocalization regime.
A corresponding generalization to several phase revolutions is provided showing a characteristic nodal pattern
among the near-degenerate eigenstates.
\end{abstract}

\maketitle

\section{Introduction}

\noindent
Symmetries play a pivotal role in physics. They codetermine or even dictate the structure
and form of interactions among the fundamental constituents of matter. Knowing the underlying
symmetries the mathematical framework of group theory \cite{Hamermesh89} provides us with a powerful toolbox which,
in particular, allows us to predict physical properties without entering detailed, often expensive,
numerical simulations. This statement holds across the disciplines of physics: in atomic physics \cite{Friedrich10}
rotational symmetry classifies the eigenstates according to their angular momentum and imposes
selection rules on their transitions whereas in solid state physics \cite{Ashcroft76,Singleton01}
the spatial periodicity of crystals yields the Bloch theorem and conservation of the crystal momentum
providing us with band structure theory. While the intimate connection between (discrete or continuous)
symmetries and constants of motion holds for the above global symmetries, the situation is much less 
transparent and explored in the case of local symmetries, i.e. a symmetries that hold only for a 
subdomain of the overall space of a given physical system. Obviously, the possibilities of having
local symmetries leads to more complex setups as compared to having corresponding global symmetries.

\noindent
Departing from e.g. crystals one well-known pathway to lower the symmetry is to replace
the strict periodicity by quasiperiodicity resulting in quasicrystals that exhibit a very special
self-similar arrangement of local but no global symmetries. Their aperiodic long-range orders places them in the huge gap 
between crystalline and disordered phases 
\cite{Macia09,Macia21,Shechtman84,Suck02,Janssen86,Berger93,Vieira05,Tanese14,Jagannathan21,Morfonios14}.
Resultingly a rich behaviour emerges such as fractal energy spectra, critical localization of eigenstates, and 
eigenstates being arranged in so-called quasibands \cite{Prunele01,Prunele02,Bandres16,Vignolo16,Macia17}.
Specifically aperiodic one-dimensional sequences exhibit a local symmetry dynamics \cite{Morfonios14}
that follows fundamental symmetry principles. However, local symmetries do not need to be arranged
quasiperiodically, which is ultimately only a very special case, but can be distributed via
emergence or design in a plethora of different ways. Fundamental for all of them is the 
fact that they show a generalization of the Bloch theorem where local symmetry-induced invariant nonlocal currents
replace the quasimomentum \cite{Kalozoumis14a}. These currents obey non-local continuity equations which
have been established and exploited for both continuous \cite{Diakonos19} and discrete \cite{Morfonios17} setups
and generalized to many-body interacting systems \cite{Schmelcher17}.

\noindent
In a number of works based on the concept of local symmetries it has been demonstrated that their presence
introduces novel properties into an otherwise unstructured system. Local symmetries can lead to
sum rules for the invariant currents causing perfectly transmitting resonances for both
quantum scattering \cite{Kalozoumis13a} and electromagnetic wave propagation through multilayer \cite{Kalozoumis13b}
dielectric media. A computational framework for wave propagation that replaces the band structure calculation for 
crystals and addresses arbitrary combinations of local symmetries has been developed and applied \cite{Zambetakis16}.
The invariant currents have been detected experimentally and used to control the wave propagation
in (lossy) acoustic waveguides \cite{Kalozoumis15} and in coupled photonic waveguide lattices \cite{Schmitt20}.
Equipping a wave scattering device with more and more local symmetries it has been shown that the corresponding
transport across this device is enhanced systematically \cite{Morfonios20}.

\noindent
A repeating observation in the analysis of locally symmetric devices is the fact that the localization
of the underlying eigenstates tends to occur on subdomains with local symmetries. This has 
consistenly been found in aperiodic long-range ordered setups \cite{Roentgen19} but even in setups
which have only isolated local symmetries \cite{Morfonios20,Schmelcher24} but are otherwise disordered.
The origin of this localization behaviour was analyzed in depth \cite{Schmelcher24} and is due to
the fact that the eigenvalue spectrum of a subdomain of a lattice is invariant w.r.t. to symmetry
transformations (reflection, translation). Therefore local symmetries imply (coupled) isospectral
subdomains and the resulting degeneracies facilitate and allow to control the (de-)localization
of the eigenstates as well as the splitting of the degenerate eigenvalues due to the couplings.
These findings have recently lead \cite{Schmelcher25} to the introduction of a new category of lattices:
the isospectrally patterned lattices (IPL). The guiding construction principle behind IPL
is the composite arrangement of isospectral (degenerate) cells that constitute
an overall lattice. An immediate set of parameters that characterizes isospectral
cells are the rotation angles of orthogonal transformations which create the cells on
basis of an underlying orthogonally transformed diagonal matrix (see following section \ref{SAH}).
IPL can therefore be specified and designed according to the changes of the phases across the lattice.

\noindent
A first recent exploration of IPL \cite{Schmelcher25} has been focusing on finite (one-dimensional) lattices
with a single phase varying according to a constant (discrete) gradient across the lattice. It has been demonstrated that
the resulting 'band' structure of this inhomogeneous setup comprises three different regimes of 
localized versus delocalized states and crossovers between them. The localized states expand
around a single center with increasing degree of excitation. The localization mechanism has been
identified to emerge as a consequence of the competition between the phase gradient and the coupling among the
isospectral cells and is therefore inherently different from other known localization mechanisms, such as Anderson
localization \cite{Abrahams10}. The relative fraction of localized vs. delocalized states in each band can be controlled
by changing the phase gradient.

\noindent
Motivated by the above findings we perform in the present work an extensive computational study of IPL
covering several setups with different phase profiles. This includes asymmetrically placed phase profiles
with no reflection symmetry and IPL with a single or several complete phase revolutions. We analyze their
spectral properties in terms of eigenvalues and eigenstates, thereby addressing their band structure and
localization properties. Specifically we discuss in section \ref{SAH} our general setup, its relevant
parameters and the appearance of the lattice Hamiltonian. To set the stage and be self-contained
we provide in Section \ref{SI} a concise summary of the features of the previously investigated symmetric IPL.
Section \ref{ASI} is then dedicated to a spectral analysis of the asymmetric IPL. Covering a complete
phase revolution is the case of investigation in section \ref{PRI} whereas several phase revolusions are
addressed in section \ref{PRIS}. Finally we conlude and provide perspectives of future research in
section \ref{PRIS}.

\section{Setup and Hamiltonian}
\label{SAH}

\noindent
Isospectrally patterned lattices consist of interconnected cells all of which share the same set
of energy eigenvalues. To ensure the latter and provide a unique parametrization of the cells they
are created by the orthogonal (or in general unitary)
transformation ${\mathbf{O}}_{{\bm{\phi}}_m}$ of a given diagonal matrix ${\mathbf{D}}$, i.e. we have for
the matrices describing the cells constituting the lattice the following appearance

\begin{equation}
{\mathbf{A}}_{m} = {\mathbf{O}}_{{\bm{\phi}}_m}^{-1} {\mathbf{D}} {\mathbf{O}}_{{\bm{\phi}}_m}
\end{equation}

\noindent
where $m \in \{1,...,N\}$ stands for the cell index and ${\bm{\phi}}_m$ represents the
set of phase angles $\{\phi^1_m,...,\phi^{N_p}_m\}$, or shortly
phases, that characterizes and parametrizes the $m-$th cell. For cells of dimension $K \times K$ 
there is $N_p=\frac{K(K-1)}{2}$ such phases. These phases ${\bm{\phi}}_m$ can in principle be chosen arbitrarily
but a properly designed IPL would, according to a first intuition, follow certain rules for the generation
of the sequences of phases with varying sites of the lattice. The enormous flexibility in choosing these sequences 
adds to the richness of IPL. Finite lattices covering certain phase intervals for each of the 
$\{\phi^1_m,...,\phi^{N_p}_m\}$ would be, pictorially speaking, inhomogeneous setups whereas periodic
or quasiperiodic setups are possible by correspondingly choosing the phase difference between neighboring
cells being rational or irrational and, principally, extending the lattice size to infinity.
The previously found degeneracy-based (de-)localization mechanisms \cite{Schmelcher24}
in symmetry-related isospectral lattice domains provides a major additional motivation
and raises expectations for novel spectral properties of the IPL. 

\noindent
Our lattice Hamiltonian therefore consists of $N$ cells and diagonal blocks ${\mathbf{A}}_{m},m \in \{1,...,N\}$
coupled via off-diagonal blocks ${\mathbf{C}}_{m}, m \in \{1,...,N-1\}$ and takes on the following appearance

\begin{eqnarray}
{\cal{H}} &=& \sum_{m=1}^{N} \left(\ket{m} \bra{m} \otimes \mathbf{A}_{m} \right)
\label{eq1} \\ \nonumber
& + & \sum_{m=1}^{N-1} \left(\ket{m+1} \bra{m} \otimes \mathbf{C}_{m} + h.c. \right) 
\end{eqnarray}

\noindent
One might therefore take the viewpoint of considering the cell subspace as internal degrees of 
freedom and the cell index as external degree of freedom. $N_s$ will in the following denote the
total number of lattice sites within and across cells. In the previous first work on IPL \cite{Schmelcher25}
as well as here we will focus on the case $K=2$ resulting in a single phase parameter $\phi$ which varies across
the cells of the lattice. Consequently we have for a grid of a single phase
${\mathbf{A}}_{m} = {\mathbf{O}}_{{\phi}_m}^{-1} {\mathbf{D}} {\mathbf{O}}_{{\phi}_m}$ or written out
explicitly this yields

\begin{equation}
\begin{aligned}
\mathbf{A}_m = &
\begin{bmatrix}
\cos \phi_m &   \sin \phi_m \\
- \sin \phi_m & \cos \phi_m \\
\end{bmatrix}
\begin{bmatrix}
d_1 &  0 \\
0 & d_2  \\
\end{bmatrix}
\begin{bmatrix}
\cos \phi_m &   -\sin \phi_m \\
\sin \phi_m & \cos \phi_m \\
\end{bmatrix}  \\
= &
\begin{bmatrix}
d_1 \cos^2 \phi_m + d_2 \sin^2 \phi_m &   (d_2 - d_1) \sin \phi_m \cos \phi_m\\
(d_2 - d_1) \sin \phi_m \cos \phi_m  & d_1 \sin^2 \phi_m + d_2 \cos^2 \phi_m \\
\end{bmatrix}
\end{aligned}
\end{equation}

\noindent
Obviously $\mathbf{A}_m$ possesses the determinant $d_1 \cdot d_2$ and the trace $d_1 + d_2$.
By substituting $\phi_m = \psi_m + \frac{\pi}{4}$ into the above expression one obtains

\begin{equation}
\begin{aligned}
\mathbf{A}_m = & \frac{d_1+d_2}{2} {\mathbb{1}} + \frac{d_2-d_1}{2}
\begin{bmatrix}
\sin( 2\psi_m) &   \cos( 2 \psi_m)\\
\cos( 2\psi_m) &   -\sin( 2 \psi_m)\\
\end{bmatrix}
\end{aligned}
\end{equation}

\noindent
which splits $\mathbf{A}_m$ into a part proportional to the identity matrix, i.e. a global offset, and a traceless contribution
comprising its nontrivial part. The coupling between the cells will be chosen as 
${\mathbf{C}}={\mathbf{C}}_m = \frac{\epsilon}{2} \left(\sigma_x + i \sigma_y \right)$ and we use
open boundary conditions for our lattice. Here $\epsilon$ denotes the coupling strength and $\sigma_i, i=x,y$
are the corresponding $2 \times 2$ hermitian and unitary Pauli matrices.
Note that the above-designed lattices represent one-dimensional configurations/chains but could easily be
generalized to higher dimensions by introducing the corresponding off-diagonal couplings.
The broad class of IPL contains several important special models. The SSH-model (see ref.\cite{Cooper19}
and references therein) is obtained
by making all phases across the lattice of equal value $\psi = \frac{\pi}{2}$ such that, after subtracting
the global offset value, the diagonal entries vanish. The dimerized character
stems then from the intracell coupling and the intercell coupling (see below).
The (time-independent) Rice-Mele model is obtained for a constant but arbitrary phase $\psi$
(see \cite{Lin20} and references therein).
We will focus in the present work on finite lattices covering a finite
interval of the phase $\phi$ and possessing a constant phase difference between neighboring cells.

\section{Symmetric IPL}
\label{SI}

\noindent
To set the stage for a prototype IPL and in order to be self-contained let us firstly recapitulate and
expand on the spectral properties of what we call a symmetric IPL which has been explored previously \cite{Schmelcher25}.
For the single phase $\phi_m$ we choose an equidistant grid of phase values centered around the specific value $\frac{\pi}{4}$.
The motivation herefore lies in the previous observation \cite{Schmelcher24} that isospectrally related cells
that gradually change across the lattice might offer the possibility to steer the localization properties of the lattice
eigenstates. The phase range covered by the lattice is then given by 
$[\frac{\pi}{4}-\frac{L}{2},\frac{\pi}{4}+\frac{L}{2}]$ with $L = \frac{\pi}{4} \cdot \frac{1}{L_f}$
where $L_f$ is a scaling factor for the phase interval covered by the lattice. For the individual
phase value of the $m-$th cell we have 
$\phi_m = \frac{\pi}{4} - \frac{L}{2} + \frac{m-1}{N-1} L, \hspace*{0.1cm} m \in \{1,...,N\}$. 
The specific center value $\phi = \frac{\pi}{4}$ is chosen because for this phase the corresponding
cell eigenvectors read $(\text{cos} \phi, - \text{sin} \phi)$ with the eigenvalue $d_1$ and 
$(\text{sin} \phi, \text{cos} \phi)$ with the eigenvalue $d_2$ which are the diagonal entries of the
matrix ${\mathbf{D}}$. Obviously, these are for $\phi = \frac{\pi}{4}$ 'maximally delocalized' whereas
$\phi = 0, \frac{\pi}{2}$ renders one of the two components of the eigenvectors zero.
Additionally, our lattice shows, by choosing for the center cell $\phi = \frac{\pi}{4}$, an inversion symmetry.

\noindent
A note is in order concerning our use of the term (de-)localized (eigen-)states. Since we address in this
work exclusively finite lattices we apply the term localized eigenstates to those states which possess
a substantial amplitude solely inside the lattice and dont reach the boundaries of the lattice or
which reach only one edge of the lattice. Delocalized
eigenstates extend over the complete lattice. As we shall see localized and delocalized states are not
intermingled but occur in specific energetical regimes and are separated 
by a finite system localization delocalization crossover (FLDC).

\begin{figure}[H]
\centering
\includegraphics[width=9cm,height=5cm]{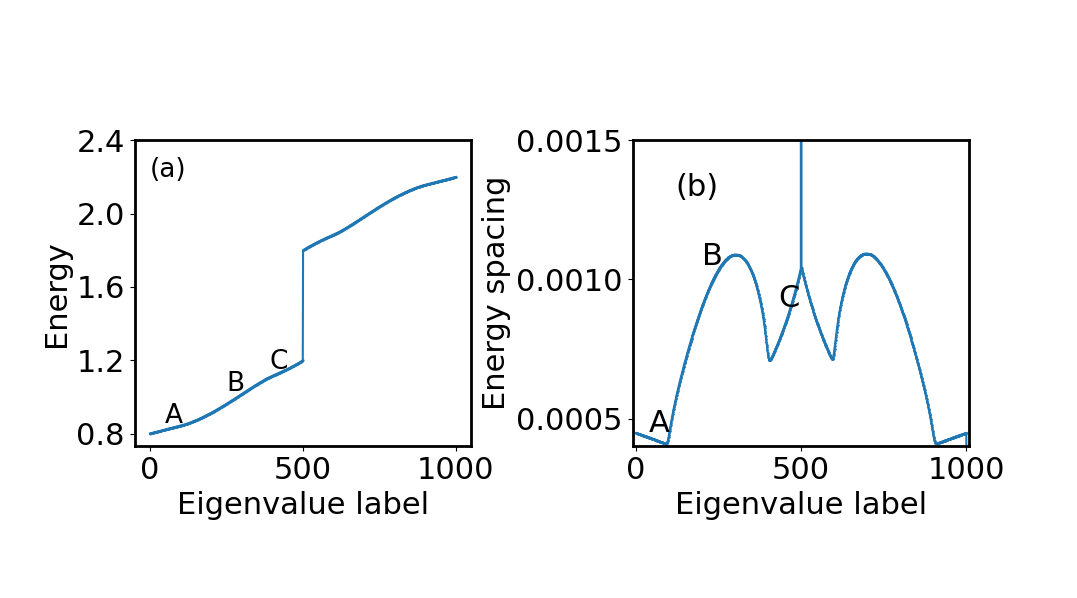}
\caption{(a) Energy eigenvalue spectrum of the equidistant $\phi$ lattice with $\frac{\pi}{8} \leq \phi \leq \frac{3\pi}{8}$,
for $d_1=1,d_2=2, L_f=1.0, \epsilon=0.2, N_s=1002$. The vertical line indicates the energy
gap between the two bands. (b) The corresponding energy eigenvalue spacing with increasing degree of excitation. 
Each band consists of three different domains labeled by A, B and C for the lowest band.
The energy level spacing shows a much more abrupt transition between those domains as compared to the
energy eigenvalue spectrum itself.}
\label{Fig:1}
\end{figure}

\noindent
Let us start our analysis by inspecting the energy eigenvalue spectrum for an IPL for an
equidistant $\phi$ lattice with $\frac{\pi}{8} \leq \phi \leq \frac{3\pi}{8}$,
for $d_2=2, d_1=2, L_f=1.0, \epsilon=0.2, N_s=1002$, see Fig.\ref{Fig:1}. Fig.\ref{Fig:1}(a)
shows the eigenvalues which are arranged in two bands separated by a large gap.
The individual bands show a behaviour which is distinctly different from the
cosine-like dispersion relation for a monomer tight-binding chain: one observes three
different energetical subdomains for each band, labeled by A,B,C for the energetically
lower band in Fig.\ref{Fig:1}, for which the slopes are different.

\begin{figure}[H]
\centering
\includegraphics[width=9cm,height=6cm]{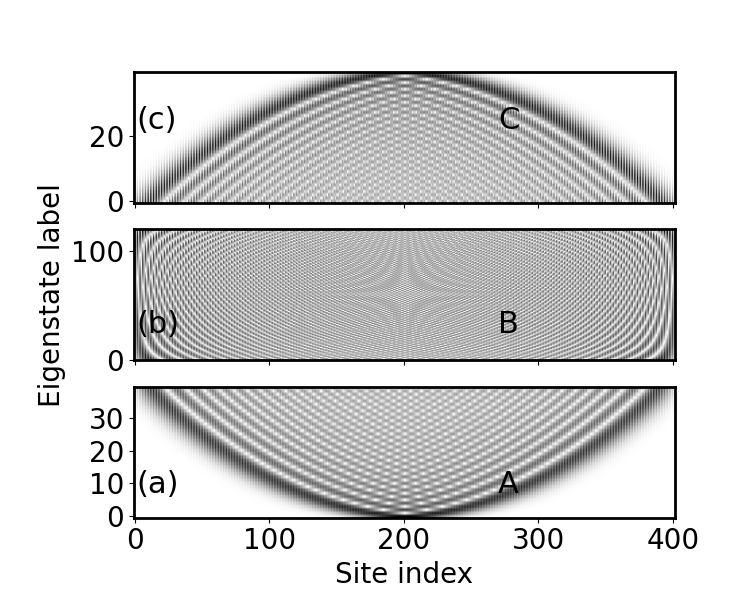}
\caption{Grey scale eigenstate map showing the absolute values of all eigenstate components for the lower
band of the spectrum of the equidistant $\phi$ lattice with $\frac{\pi}{8} \leq \phi \leq \frac{3\pi}{8}$,
i.e. placed symmetrically w.r.t. $\frac{\pi}{4}$, for $d_1=1,d_2=2, L_f=1.0, \epsilon=0.2, N_s=402$.  
The grey scale has been renormalized for each eigenstate row. 
The labeling of the eigenstates (vertical axis) is reset to zero within each energy domain with a total of $201$ states. 
(a,b,c) show the eigenstate profiles according to the three energy domains A,B,C of the lower band in Fig. \ref{Fig:1}; 
domains A and C show centered and localized eigenstates whereas domain B consists of eigenstates that extend over
the complete lattice and reach out to its boundaries.}
\label{Fig:2}
\end{figure}

\noindent
They indicate a qualitatively
different behaviour of the level structure for these different subdomains.
This becomes
more pronounced when inspecting the energy eigenvalue spacing instead of the absolute eigenvalues,
as can be seen in Fig.\ref{Fig:1}(b). The energy spacing with increasing degree of excitation
shows a very different behaviour in the three subdomains A,B,C. 
In subdomain A the spacing is slightly decreasing with increasing
degree of excitation but is in zeroth order approximation constant.
In subdomain B the spacing behaves highly nonlinear: it first increases in an approximately parabolic manner and then, after 
reaching a maximum, it decreases correspondingly ending at a point which is at a higher energy
compared to the beginning of the subdomain B. Finally in regime C the energy spacing increases
strongly in a slightly nonlinear way, meaning that a linear fit represent a moderately good approximation
in this regime. This behaviour repeats in an inverted fashion in the second band.

\noindent
The question which arises in view of the characteristic eigenvalue spectrum shown in Fig.\ref{Fig:1} addresses the
origin of the observed spectral behaviour. To illuminate this we show in Fig.\ref{Fig:2} 
a grey scale eigenstate map presenting the absolute values of all eigenstate components for all eigenstates of the lower
band of the spectrum for an equidistant $\phi$ lattice with $\frac{\pi}{8} \leq \phi \leq \frac{3\pi}{8}$
for the same parameters as in Fig.\ref{Fig:1} except for the value $N_s=402$. This map is partitioned according
to the subdomains A,B,C of Fig.\ref{Fig:1} in the subfigures Fig.\ref{Fig:2}(a,b,c). We observe that regions
A and C correspond to localized eigenstates which are centered around the center cell with $\phi = \frac{\pi}{4}$.
While the global ground state shows an envelope behaviour of approximately Gaussian character \cite{Schmelcher24}
we observe (not shown here) with increasing degree of excitation an increasing number of nodes of this envelope.
The localization mechanism \cite{Schmelcher24} involves the competition between the phase gradient across the lattice
and the coupling among the cells. Emerging from the center cell the localized eigenstates spread with increasing
degree of excitation until they reach the boundary of the lattice for the eigenstate at the crossover from
subdomain A to B or from subdomain B to C in the reverse manner. In contrast to this subdomain B consists of
delocalized eigenstates that extend over the complete lattice and reach to its boundaries.

\begin{figure}[H]
\centering
\includegraphics[width=9cm,height=8cm]{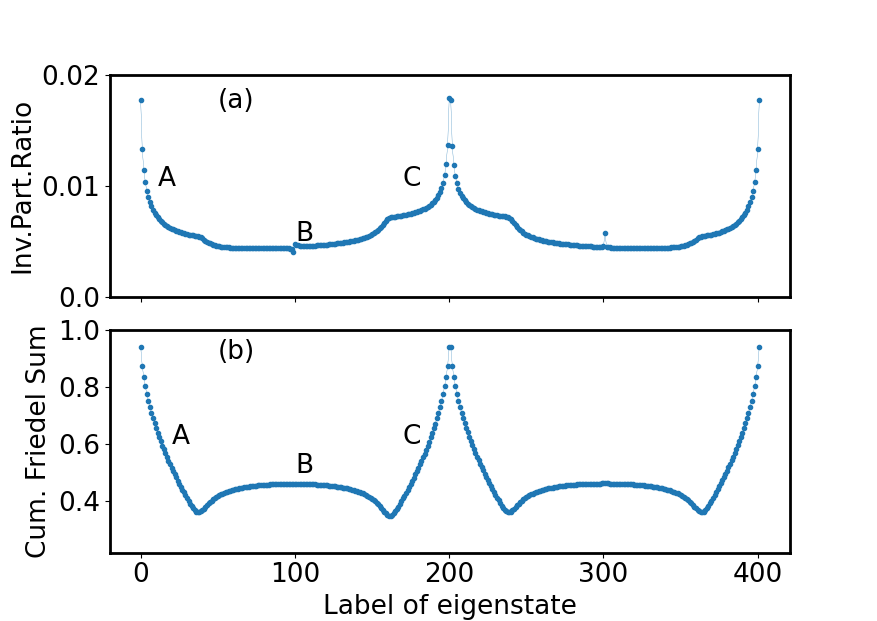}
\caption{(a) The inverse participation ratio for all eigenstates across both bands. Parameters are the
same as in Fig. \ref{Fig:2}. (b) The corresponding cumulative Friedel sum. Both measures for localization
reflect the domains A,B,C: eigenstates in A,C posssess typically higher values and rapidly change with
varying degree of excitation for both the inverse participation
ratio and the cumulative Friedel sum whereas eigenstates in B form (apart from the corresponding transition region)
a flat valley of low values.}
\label{Fig:3}
\end{figure}

\noindent
To quantify the above localization behaviour of the eigenstates of the IPL we determine their inverse participation
ratio which is defined as $r = \sum_{i=1}^{N} |\psi_i|^4 \in [N^{-1},1]$, where $\psi_i$ is the 
eigenvector component on site $i$ of a given eigenstate $\psi$. The IPR of an eigenstate possesses the maximal
value one for an eigenstate localized on a single site of the lattice and it has the minimal value $\frac{1}{N}$ for 
a state uniformly extended over the lattice. Fig.\ref{Fig:3} shows the IPR for all eigenstates of the lowest band
shown in Fig.\ref{Fig:2}. Starting from the ground state we observe a decrease of the IPR with increasing degree
of excitation in the subdomain A until the crossover to subdomain B takes place and within the latter a plateau-like
behaviour of the IPR occurs. Finally, in subdomain C, the IPR raises again with further increasing degree of 
excitation, reflecting the increasing degree of localization culminating again at the upper band edge into a Gaussian-like
envelope behaviour of the corresponding eigenstate. Therefore, the IPR clearly signalizes the two FLDC occuring in each
band. 

\noindent 
The IPR as a localization measure does not depend on the position at which a state is
localized within a given lattice system. It is, however, largely insensitive to the spatial 
state profiles \cite{Gong16}. 
To address this issue, and exploit an alternative localization measure, we employ
the cumulative Friedel sum (CFS) of a given state \cite{Morfonios20}.
The CFS is based on a measure developed in \cite{Gong16} which is inspired by the Friedel
sum rule \cite{Langreth66}. 
It reflects more details of the spatial profile by using its cumulative sum
$P_n = \sum_{m=1}^n |\psi_m|^2$ up to site $n$ and reads

\begin{equation} \label{eq:cfs}
 f = \frac{1}{2N} \left|\sum_{n=1}^N \left(e^{2\pi i P_n} + 1\right)  \right| ~~\in [N^{-1},1]
\end{equation}

\noindent
Larger (smaller) values for the CFS indicates a more (less) localized state, 
though now taking into account its total spatial extent instead of only its 
site participation. Inspecting the CFS for the eigenstates in Fig.\ref{Fig:3}(b)
a similar behaviour to that of the IPR can be observed: in the subdomain A, starting
with the ground state in the band, we have a strong decrease of the CFS values. In subdomain
B there is a minor variation of the CFS values whereas in subdomain C a strong increase is 
obtained for an increasing degree of excitation. All of this behaviour reflects again the
fact that the eigenstates in the subdomains A,C correspond to localized states that gradually
delocalize and vice versa, and that the eigenstates in subdomain B are all delocalized. It should
be noted that the absolute values of the CFS as well as their variations are much larger compared to those of the IPR.

\begin{figure}[H]
\centering
\includegraphics[width=9cm,height=7cm]{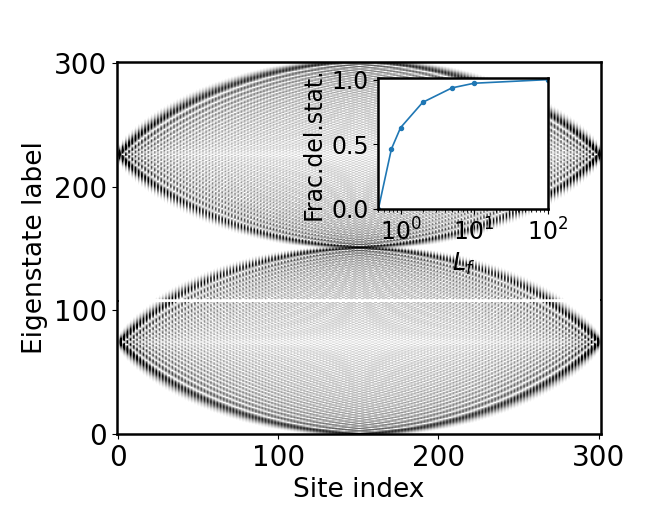}
\caption{Grey scale eigenstate map showing the absolute values of all eigenstate components
for the equidistant $\phi$ lattice with $0 \leq \phi \leq \frac{\pi}{2}$,
i.e. placed symmetrically w.r.t. $\frac{\pi}{4}$, for $d_1=1,d_2=2, L_f=0.5, \epsilon=0.3, N_s=302$.
Almost all eigenstates are localized in this case. The inset shows the fraction of delocalized states 
with varying parameter $L_f$ which relates to the corresponding phase gradient. Here $N_s = 1002, \epsilon=0.2$.
This demonstrates that the fraction of (de-)localized states can be varied from no extended states to
extended states only.}
\label{Fig:4}
\end{figure}

\noindent
Let us now inspect how the above-observed behaviour of a divided eigenstate space consisting of localized and
delocalized states changes if we change the relevant parameters. For the above-considered case we had a parameter
value $L_f=1.0$ corresponding to a total phase interval $\frac{\pi}{8} \leq \phi \leq \frac{3\pi}{8}$. Let us now
focus on the case $L_f = 0.5$ for which $0 \leq \phi \leq \frac{\pi}{4}$. Fig.\ref{Fig:4} shows the 
eigenstate map for this parameter value, which demonstrates that almost all states are localized and subdomains A
and B have taken over the eigenstate profile map where subdomain C consists only of a very few 'limiting' states.
This is indicative of the fact that, with varying parameter $L_f$, and consequently with varying phase gradient
the fraction of localized versus delocalized eigenstates can be varied arbitrarily. Indeed, the inset of
Fig.\ref{Fig:4} shows that this fraction can be tuned continuously from the value zero to one while varying
$L_f$ from $0.5$ to very large values. It should be noted that for $L_f \rightarrow \infty$ our IPL becomes
periodic since the phase interval 'collapses' to a single value and in this case only extended states survive.

\begin{figure}[H]
\centering
\includegraphics[width=9cm,height=6cm]{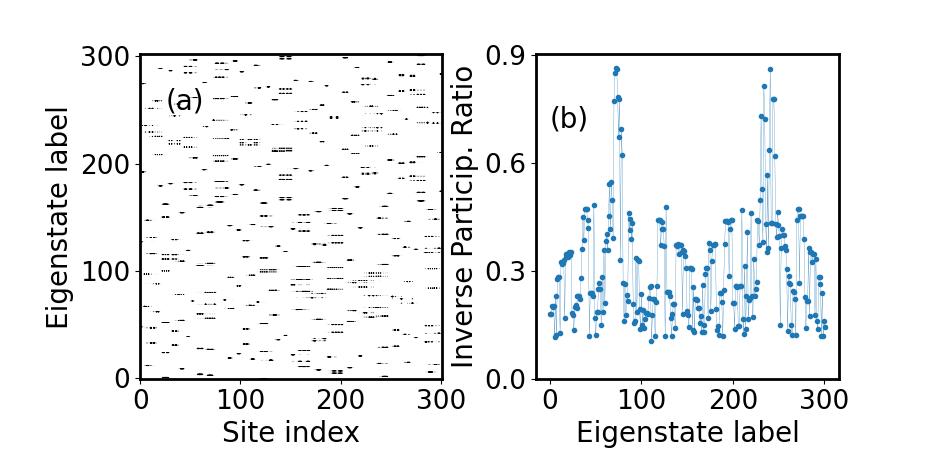}
\caption{(a) Grey scale eigenstate map showing the absolute values of all eigenstate components
for a lattice with randomly chosen on-site energies $d_1=1,d_2=2$ and a coupling strength $\epsilon=0.2$ for $N_s=302$.
Inspection by eye shows already that the spreading of the eigenstates is overall by orders of magnitude smaller
than the lattice size. (b) The corresponding inverse participation ratio for all eigenstates is irregularly
oscillating opposite to the smooth behaviour observed for the isospectrally patterned lattice,
see Fig.\ref{Fig:3}. It is by orders of magnitude larger due to the strong localization.}
\label{Fig:5}
\end{figure}

\noindent
It is instructive to compare the above results on the eigenstate localization properties with examples
of random distributions. To this end Fig.\ref{Fig:5}(a) shows the eigenstate map for a lattice
with a random sequence of on-site energies for the values one and two, but otherwise the same parameter
values as in Fig.\ref{Fig:2} except for
$N_s = 302$. There is two obvious major differences of the localization behaviour of the eigenstates compared to
our IPL (see Fig.\ref{Fig:2}). First, the eigenstates do not possess a single center as it is the case
for the IPL, but are, as expected, localized around many different centers reflecting the random design with many 'impurities'.
Second, the localization length of those eigenstate profiles is much smaller compared to the
deterministic localization length of the IPL eigenstates which is a 'cooperative' effect of the phase gradient
and the coupling strength. To quantify the localization behaviour Fig.\ref{Fig:5}(b) shows the corresponding
IPR which exhibits irregular fluctuations on a scale much larger than the typical values occuring
for the IPL (see Fig.\ref{Fig:3}). This is, as mentioned-above, due to the stronger localization in the 
random case. While the IPR for the IPL varies smoothly with increasing degree of excitation, the IPR of
our random on-site energy case shows no such smooth behaviour.

\begin{figure}[H]
\centering
\includegraphics[width=9cm,height=6cm]{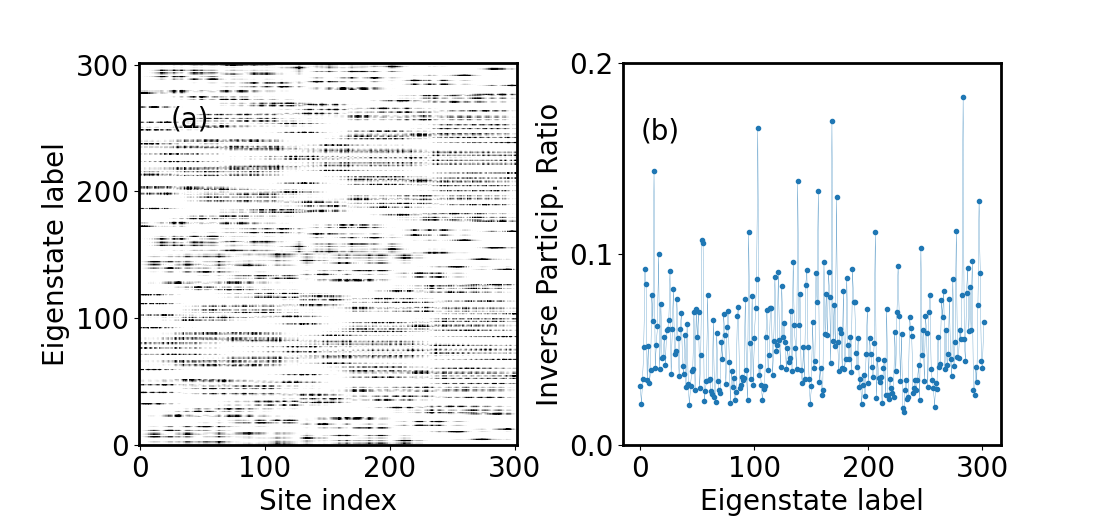}
\caption{(a) Grey scale eigenstate map showing the absolute values of all eigenstate components
for a lattice with randomly chosen phases in the range $\frac{\pi}{8} \leq \phi \leq \frac{3\pi}{8}$
and $d_1=1,d_2=2$ with a coupling strength $\epsilon=0.2$ for $N_s=302$.
(b) The corresponding inverse participation ratio for all eigenstates. For this phase disorder case, the
eigenstates are much more spreading out as compared to the random on-site energy case, see Fig.\ref{Fig:5}, also
reflected in the shown smaller values of the inverse participation ratio.}
\label{Fig:6}
\end{figure}

\noindent
As a second case of comparison we show in Fig.\ref{Fig:6}(a) the eigenstate map for an IPL with randomly
chosen phases in the range $\frac{\pi}{8} \leq \phi \leq \frac{3\pi}{8}$. Also here we observe that there
is no smooth structures and crossover between localized and delocalized states as for the symmetric IPL
discussed above. The band(s) consist exclusively of differently centered localized eigenstates. However, a 
comparison with Fig.\ref{Fig:5}(a) reveals that the localization length is typically larger compared to the
random on-site energy case. Indeed, Fig.\ref{Fig:6}(b) shows that the irregularly fluctuating values of the 
IPR are consistently lower for the case of phase randomization, but still significantly larger compared to
our original symmetric IPL (see Fig.\ref{Fig:3}(a)).

\noindent
Finally, a note is in order concerning the impact of the (de-)localized state decomposition of the IPL
upon contacting it from the outside, i.e. probing the IPL transport properties. It is to be expected
that for the energetical subdomains A,C of localized eigenstates the transport would be strongly suppressed
whereas for the delocalized eigenstate subdomain B transport takes place. Therefore, the IPL would allow us to
controllably provide transport in an energetical window determined by the width of the subdomain of delocalized
states, which is tunable. A detailed investigation of the transport properties of IPL goes however beyond the scope of the
present investigation which focuses on its spectral properties.

\begin{figure}[H]
\centering
\includegraphics[width=9cm,height=7cm]{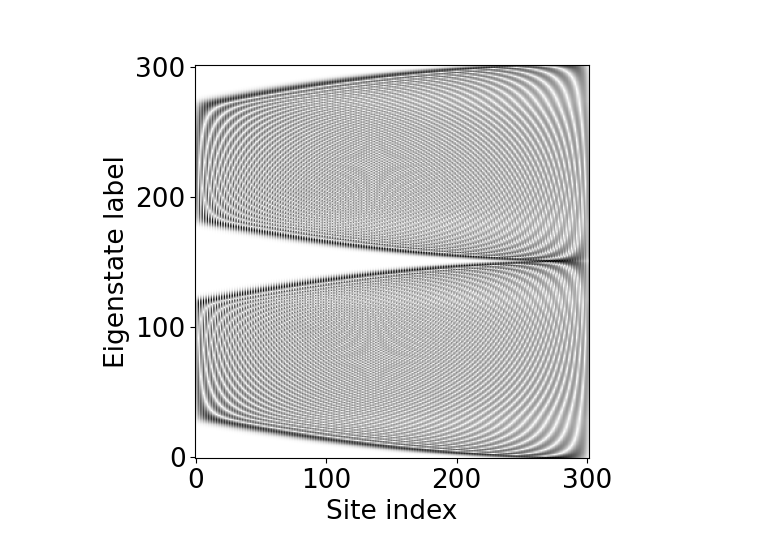}
\caption{Grey scale eigenstate map showing the absolute values of all eigenstate components for both
bands of the spectrum of the equidistant $\phi$ lattice with $\frac{\pi}{8} \leq \phi \leq \frac{\pi}{4}$,
i.e. placed asymmetrically w.r.t. $\frac{\pi}{4}$, for $d_1=1,d_2=2, L_f=2.0, \epsilon=0.3, N_s=302$.  
The grey scale has been renormalized for each eigenstate row. Eigenstates are now centered around the right
edge of the lattice and gradually extending towards the left edge with increasing (decreasing) degree of
excitation from the lower (upper) band edge. In between is a regime of delocalized states extending over the
complete lattice.}
\label{Fig:7}
\end{figure}

\section{Asymmetric IPL}
\label{ASI}

\noindent
There is many different possibilities for choosing the sequence of phases underlying an IPL.
In the previous section we have chosen an equidistant phase grid covering a finite total phase interval
$[\frac{\pi}{4} (1 - \frac{1}{2 L_f}),\frac{\pi}{4} (1 + \frac{1}{2 L_f}]$ where the
phase distribution is placed symmetrically around its center value $\phi = \frac{\pi}{4}$.
As a natural first extension of this symmetric IPL we will now break this inversion symmetry
and place the phase interval covered by the lattice asymmetrically around the value $\frac{\pi}{4}$. 

\noindent
Fig.\ref{Fig:7} shows the eigenstate map for the completely asymmetric case with 
$\frac{\pi}{8} \leq \phi \leq \frac{\pi}{4}$ where $\frac{\pi}{4}$ occurs at the edge of the
lattice. We observe that the ground state is now localized at the right edge of the lattice
and, with increasing degree of excitation, the eigenstates extend increasingly into the interior of the lattice,
thereby always reaching out to the right edge of the lattice. This constitutes the first subdomain
of each band. The second subdomain consists of delocalized states extending over the complete lattice
whereas the third subdomain possesses the reverse structure of the first one. We therefore encounter,
similar to the symmetric IPL, also two FLDC for the asymmetric IPL. Choosing the phase interval
in between the symmetric and completely asymmetric case (not shown here) will shift the location/centering of the 
ground state and consequently of all following excited localized states of the corresponding first
subdomain continuously from the center of the lattice to its edge. This tunability exists equally for
the case of a left edge-based localized eigenstate domain simply by choosing for the completely
asymmetric case the phase interval $\frac{\pi}{4} \leq \phi \leq \frac{3\pi}{8}$. In a nutshell,
breaking the inversion symmetry w.r.t. $\frac{\pi}{4}$ of the covered phase interval of the lattice 
allows to continuously shift the centering of the localized eigenstate subdomains thereby leading to
a tunable asymmetry.

\begin{figure}[H]
\centering
\includegraphics[width=9cm,height=7cm]{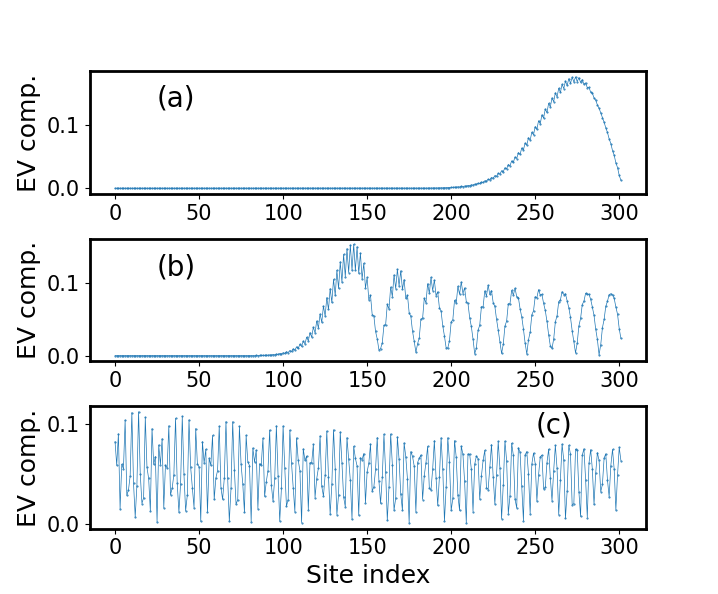}
\caption{Individual eigenstate profiles (magnitude of the eigenvector components)
for the ground state, the $10$th and $70$th eigenstate in subfigures (a,b,c), respectively.
Equidistant $\phi$ lattice with $\frac{\pi}{8} \leq \phi \leq \frac{\pi}{4}$,
i.e. placed asymmetrically w.r.t. $\frac{\pi}{4}$, for $d_1=1,d_2=2, L_f=2.0, \epsilon=0.3, N_s=302$.
The asymmetric oscillatory structure extending over (parts of) the lattice is clearly visible.} 
\label{Fig:8}
\end{figure}

\noindent
Fig.\ref{Fig:8} shows the eigenstate profiles for the ground state,
the $10$th and $70$th eigenstate in subfigures (a,b,c), respectively,
for the equidistant $\phi$ lattice with $\frac{\pi}{8} \leq \phi \leq \frac{\pi}{4}$, corresponding
to Fig.\ref{Fig:7}. The ground state in Fig.\ref{Fig:8}(a) is, as discussed above, localized next to the right boundary
of the lattice and possesses an asymmetric profile around its maximum with a tail reaching into
the inner part of the lattice. The $10$th eigenstate (see Fig.\ref{Fig:8}(b)) shows a strongly regularly oscillating
behaviour emerging from the right edge of the lattice and covering slightly more than half of the lattice size.
This state, obviously, still belongs to the subdomain of localized states near the lower band edge.
Finally Fig.\ref{Fig:8}(c) shows the $70$th eigenstate which exhibits high frequency oscillations and some
beating behaviour with a significantly lower frequency: it extends over the complete lattice and therefore
belongs to the subdomain of delocalized states.

\noindent
Let us contemplate about the impact of the asymmetric localized states particularly in the
above-shown case of complete asymmetry. If we contact this asymmetric IPL via left and right suitably
matched leads it can be conjectured that the above-discussed right centered localized states (subdomain A)
would allow for a penetration
of the eigenstates (of the IPL plus leads) that emerge from the right into the IPL with some tunable and energy-dependent
depth. For the same energies in the subdomain of localized states, a penetration from the left lead would be
strongly suppressed. For higher energies (subdomain B) transport across the IPL would take place and would be followed
(subdomain C) by an inverted scenario as compared to subdomain A.

\begin{figure}[H]
\centering
\includegraphics[width=9cm,height=5.3cm]{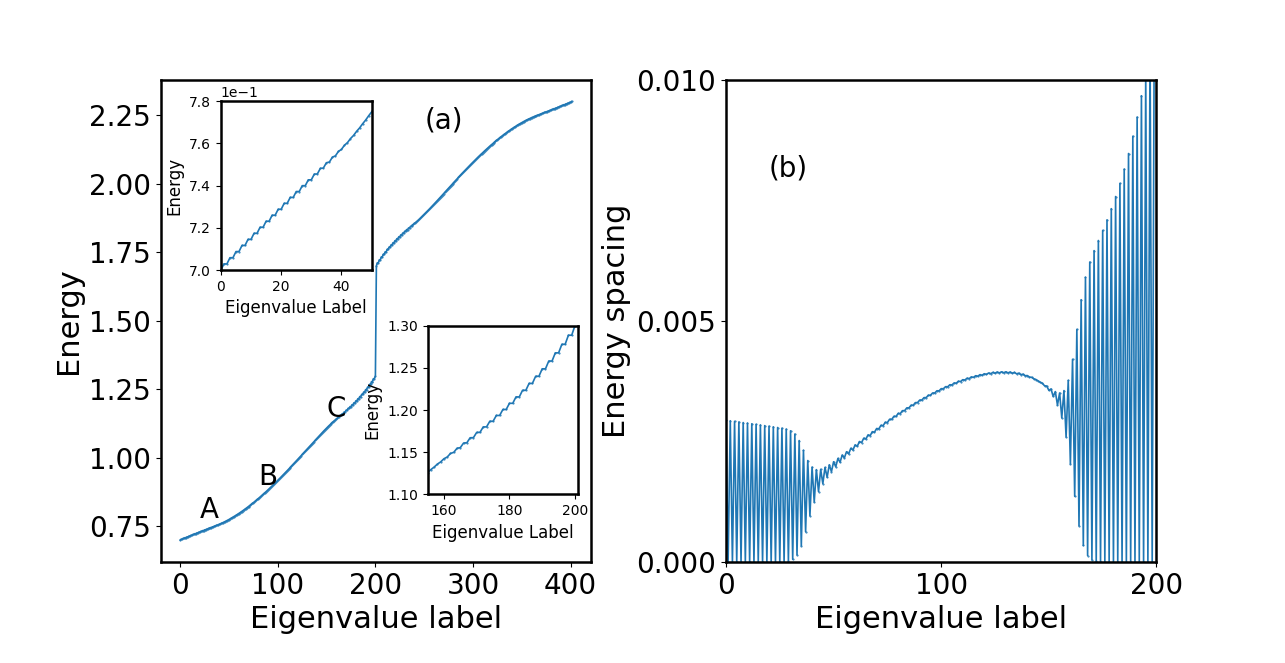}
\caption{(a) Energy eigenvalue spectrum of the equidistant $\phi$ lattice with one complete oscillation within
the interval $\frac{\pi}{8} \leq \phi \leq \frac{3\pi}{8}$, for $d_1=1,d_2=2, L_f=1.0, \epsilon=0.3, N_s=402$. 
A, B and C indicate the different domains in the corresponding bands. The insets show corresponding 
magnifications for the regimes at the lower (upper inset) and upper (lower inset)
band edge of the lower band showing a plateau-like structure. (b) The energy eigenvalue
spacing spectrum for the first band. The latter shows that the energy levels are for low (high) energies in the band
approximately pairwise degenerate whereas in the central part of the band this (approximate) degeneracy is lifted.}
\label{Fig:9}
\end{figure}

\section{IPL with a single phase revolution}
\label{PRI}

\noindent
Let us now explore the case for which the phase variation across the finite lattice is such that it increases
first and decreases thereafter. We therefore consider an oscillation of the phase covering a phase
interval $\frac{\pi}{8} \leq \phi \leq \frac{3\pi}{8}$ first by increasing the phase and, upon reaching
the turning point $\frac{3\pi}{8}$, by decreasing the phase with a constant phase difference from cell to cell.
Such an oscillatory behaviour includes within each half oscillation the point of inversion symmetry
$\frac{\pi}{4}$ for the symmetric IPL. Fig.\ref{Fig:9}(a) shows the energy eigenvalue spectrum with increasing
degree of excitation. Also in this case three different distinct subdomains can be identified for each
band. A magnification of the regions around the lower (upper inset) and upper (lower inset) band edge are shown as insets.
Here a plateau-like structure can be observed which is more pronounced the closer we are energetically to the corresponding
band edge. This structure dissolves as we move closer in energy towards the center of the band.
Upon inspection of the energy eigenvalue spacing, see Fig.\ref{Fig:9}(b), we observe a near zero value for every second
spacing: the energy eigenvalues arrange in near degenerate pairs, being closer to degeneracy the
closer the energy is to the band edge. It is important to note that the ground state is non-degenerate
(hardly visible in Fig.\ref{Fig:9}(a,b) due to the finite resolution) and the pairing starts only
with the excited states.

\noindent
Moving with the energy towards the center of the band the near degeneracies are lifted 
beyond the $\approx 40$-th eigenstate (see Fig.\ref{Fig:9}(b)) and the energy spacing
exhibits an arc-like behaviour of non-degenerate eigenvalues. The latter structure occupies a large central part of the 
spectrum. It is important to note that the extension of the subdomains A,B,C of localized vs. delocalized states
does not coincide with the regimes for which near degenerate pairs occur or not: near degeneracy is encountered
only for localized states sufficiently far from the boundaries of the IPL. A few remarks are in order and
are illuminating w.r.t. the observed spectral features: let us address some simplified models 
which are more or less different from the IPL but possess all the property that they are lattices of coupled isospectral
cells with a single complete phase revolution. Our first simplified model goes as follows. Taking as a first half
the IPL for e.g. $\frac{\pi}{8} \leq \phi \leq \frac{3\pi}{8}$ and augmenting this lattice by a decoupled second
half lattice which is the inverted of the first one, it goes without saying that we will obtain a spectrum
exclusively consisting of (exactly) doubly degenerate pairs of eigenvalues,
due to the isospectrality of the (decoupled) two halves of the lattice.
It should be noted that our IPL for a single complete phase revolusion is not inversion symmetric
around $\frac{3\pi}{8}$ due to the 'non-inverted' phase-based cells in the course of the construction
of the IPL. For our second model we use the first model and introduce a coupling between the two halves of it.
We then observe the formation of near degenerate pairs close to the band edges as described above for the IPL.
This two-fold near degeneracy holds also for the ground state, opposite to what we have observed above for
the IPL.  Further differences between this second model and the IPL reveal themselves only upon
inspecting the eigenstate profiles, which we will accomplish as a next step.

\begin{figure}[H]
\centering
\includegraphics[width=9cm,height=7cm]{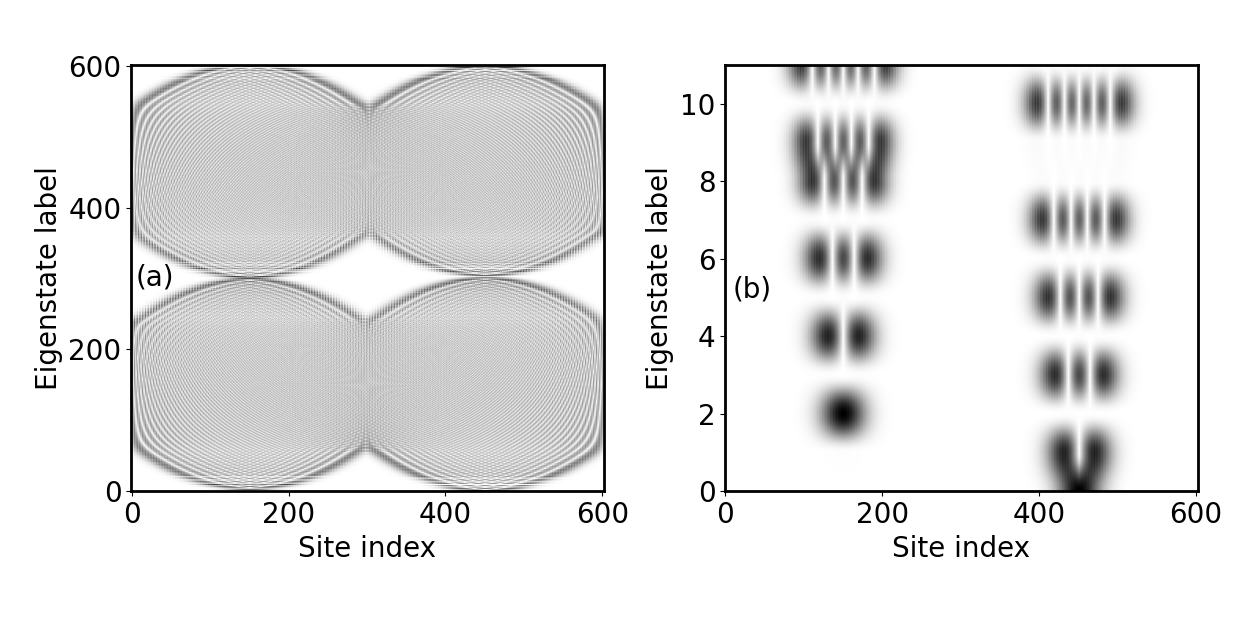}
\caption{(a) Grey scale eigenstate map showing the absolute values of all eigenstate components for both
bands of the spectrum of the equidistant $\phi$ lattice with one complete oscillation within the interval
$\frac{\pi}{8} \leq \phi \leq \frac{3\pi}{8}$ for $d_1=1,d_2=2, \epsilon=0.3, N_s=602$. The grey scale has
been renormalized for each eigenstate row. We observe two centers from which localized eigenstates spread
(lower band edge) with the reverse process, i.e. focusing, happening at the upper band edge.
(b) Magnification of the grey scale eigenstate map for the twelve energetically lowest eigenstates.
The nodeless localized states appear spatially isolated at different energies in the first and second half of 
the (finite) lattice. Near degenerate and spatially separated eigenstates (such as the states with eigenvalue
label one and two) belonging to the first and second half of the lattice possess a different nodal structure.}
\label{Fig:10}
\end{figure}

\noindent
Fig.\ref{Fig:10}(a) shows the eigenstate map for the IPL for both bands of the spectrum for an
equidistant $\phi$ lattice with one complete oscillation within the interval 
$\frac{\pi}{8} \leq \phi \leq \frac{3\pi}{8}$. Now two centered series of localized states are 
observed, each emerging and spreading around the symmetry point $\phi = \frac{\pi}{4}$ which is met
twice in the course of a single phase revolution. Upon increasing the energy these localized states
meet the boundaries of the IPL where an FLDC is encountered. As a consequence, with further increasing
energy, a subdomain of states delocalized across the complete IPL occurs. Following on this subdomain
localization takes over again culminating in a state at the upper band edge of the lower band which
consists again of two well-isolated centered distributions around $\phi = \frac{\pi}{4}$.
This behaviour repeats for the upper band.

\noindent
Fig.\ref{Fig:10}(b) allows for a closer inspection of the profiles of the energetically lowest states.
As we know from Fig.\ref{Fig:9}(a,b) we have energetically near degenerate pairs of eigenstates in the
low energy and high energy part of the spectrum and a ground state which is non-degenerate.
Surprisingly Fig.\ref{Fig:10}(b) now shows that these near degenerate pairs consist of well-localized
eigenstates each centered around $\phi = \frac{\pi}{4}$ in the first/second half of the oscillation
of the phase, and, importantly, with a different nodal structure i.e. they belong to a different
neighboring degree of excitation. This is to be contrasted with the features of the second simplified
model (coupled inversion symmetric setup) addressed above showing a near degenerate pairing of
localized eigenstates including the ground state and in particular showing for the eigenstates of a
given pair the same degree of excitation i.e. nodal structure.

\begin{figure}[H]
\centering
\includegraphics[width=9cm,height=6cm]{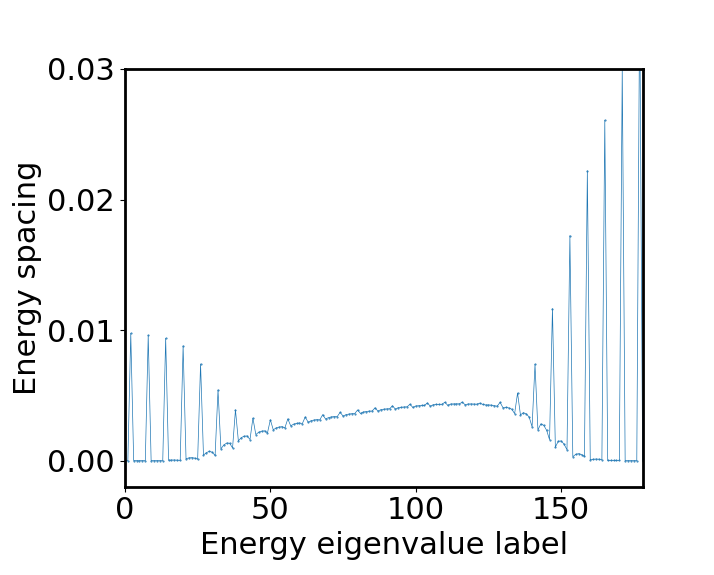}
\caption{Energy eigenvalue spacing spectrum of the equidistant $\phi$ lattice with
three complete oscillations within the interval $\frac{\pi}{8} \leq \phi \leq \frac{3\pi}{8}$,
for $d_1=1,d_2=2, L_f=1.0, \epsilon=0.3, N_s=362$. A six-fold near degeneracy is observed for energies
close to the lower (upper) band edge which is lifted in the center of the band.} 
\label{Fig:11}
\end{figure}

\section{IPL with several phase revolutions}
\label{PRIS}

\noindent
Finally we focus on an IPL with several phase oscillations taking place in the interval
$\frac{\pi}{8} \leq \phi \leq \frac{3\pi}{8}$. Fig.\ref{Fig:11} shows the eigenvalue spacing
spectrum with increasing degree of excitation for the case of three complete oscillations and the
total number of sites being $N_s=362$ for the energetically lower band.

\begin{figure}[H]
\centering
\includegraphics[width=9cm,height=5.5cm]{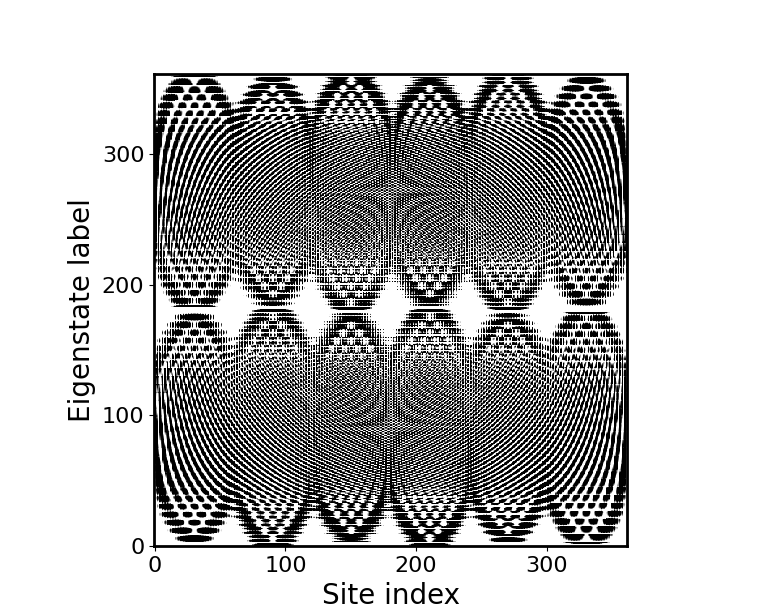}
\caption{Grey scale eigenstate map showing the absolute values of all eigenstate components for both
bands of the spectrum of the equidistant $\phi$ lattice with three complete oscillations within the interval
$\frac{\pi}{8} \leq \phi \leq \frac{3\pi}{8}$ for $d_1=1,d_2=2, \epsilon=0.3, N_s=362$. The vertically displaced
sequences of eigenstate profiles for the localized eigenstates in the first and second halves of the lattice oscillations
is visible upon close inspection.} 
\label{Fig:12}
\end{figure}

\begin{figure}[H]
\centering
\includegraphics[width=9cm,height=8cm]{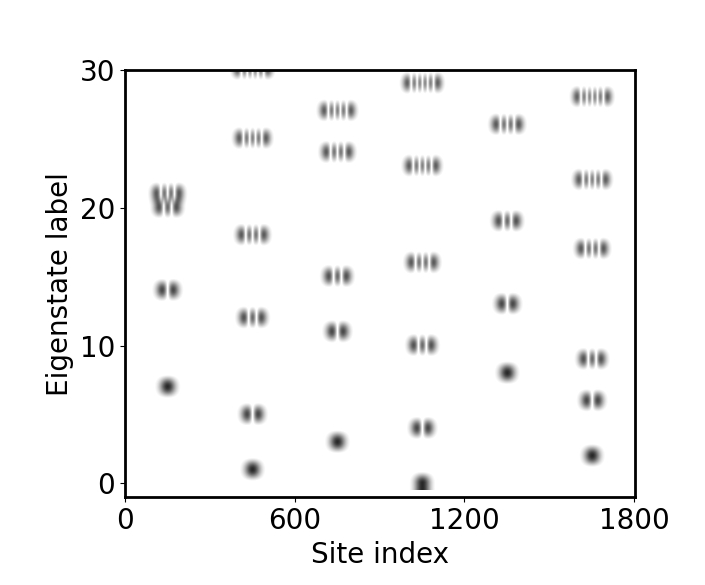}
\caption{Grey scale eigenstate map showing the absolute values of all eigenstate components for energy eigenstates
close to the lower bands band edge for the equidistant $\phi$ lattice with three complete oscillations within the interval
$\frac{\pi}{8} \leq \phi \leq \frac{3\pi}{8}$. Parameters are $d_1=1,d_2=2, \epsilon=0.3, N_s=1802$.} 
\label{Fig:13}
\end{figure}

\noindent
The isolated peaks between
multiple near zero spacing values indicate that there exist multiplets of near degenerate eigenstates
separated by 'gaps'. Indeed, a closer inspection reveals that the ground state is three-fold near degenerate 
whereas all excited states are six-fold near degenerate. This statement holds for energies in the low
energy domain whereas beyond the $\approx 50$-th eigenstate this near degenerate character is lost and
the center of the band is constituted by energetically well-separated eigenstates, similarly to what
has been observed in section \ref{PRI} for an IPL with a single oscillation.

\noindent
Finally, approaching the
upper band edge the reverse process of near degenerate multiplet formation is observed. The energy spacing
between the near degenerate multiplets is much larger and varies more strongly with increasing degree
of excitation for the eigenstates close to the upper band edge as compared to those close to the lower
band edge.

\noindent
Fig.\ref{Fig:12} shows the corresponding eigenstate map which, as expected, shows subdomains of localized
and delocalized eigenstates in each band. Now, however, we have a six-fold centering of the well-separated eigenstates
in the subdomain of localized eigenstates and an increasing spreading of them with increasing degree of excitation
finally culminating in the corresponding FLDC. Three near-degenerate states including the (global) ground state
can be identified. This becomes even more transparent and evident upon inspecting Fig.\ref{Fig:13} which shows
the eigenstate map for the low energy part of the first band. Here also the different degree of excitation and
nodal structure occuring pairwise in the six-fold near degenerate multiplets can be observed. For reasons of
illustration we have here chosen $N_s = 1802$.

\section{Conclusions and Outlook}
\label{CAO}

\noindent
Isospectrally patterned lattices are composed of cells with the same spectral content i.e. possessing the
same eigenvalue spectrum. This spectral degeneracy of the isolated cells is a key ingredient in order to
steer the localization properties of the lattice eigenstates once a coupling between the cells in the lattice is turned on. 
What makes the IPL easily accessible and straightforward to design is their parametrization via the phases/angles.
Different cells of the same spectral content are connected via an orthogonal or unitary transformation for a 
given set of transformation angles. The overall lattice can then be systematically designed in many different ways:
there is many possible choices of the variation of the angles across the lattice. In a first work \cite{Schmelcher25}
the focus was on an equidistant phase lattice covering a finite interval centered around an inversion center $\frac{\pi}{4}$.
The resulting 'band' structure consists two bands each comprising three different regimes of localized and delocalized
states for the inherently finite lattice. The localization mechanism is based on a competition of the phase gradient
and the coupling among the cells and leads to a single center localization behaviour with a characteristic
length scale and a systematic increased spreading around this center with an increasing degree of excitation. Varying the
phase gradient allows for a tuning of the fraction of localized versus delocalized states from only localized to
exclusively delocalized states.

\noindent
In the present computational study of the IPL we have detailed their properties and have, in particular, explored
the spectral properties of several novel setups. Among these is, in the first instance, an equidistant phase lattice
which is centered asymmetrically around the distinct value $\frac{\pi}{4}$. We observe that the eigenstates are now
also decentered and can be shifted continuously from the center of the lattice to its edges. Left and right localization
of the eigenstates allow for a controllable depth of 'penetration' of them into the interior of the lattice.
In a second step we investigated a phase lattice with a complete revolution of the phase. This setup leads to some
peculiar spectral structures. While the three subdomains of each band still persist their spectral content has
changed substantially. Importantly, now the single center localization
of the eigenstates is replaced by a two center localization behaviour. 
Close to the band edges a pairwise near degeneracy is encountered which lifts with increasing
degree of excitation of the localized eigenstates and finally disappears completely in the regime of delocalized
eigenstates. The reverse process is found close to the upper band edge. Notably the ground state is non-degenerate
in this scenario and the near degenerate pairs involve localized eigenstates with a different nodal structure.
This generalizes to the case of several phase revolutions which has also been analyzed in the present work.

\noindent
There is several open directions of investigation for the IPL in the future. One of them is based on
the observation that with increasing coupling among the cells the band gap closes and opens again -
here the questions is whether this is indeed accompanied by a topologically nontrivial behaviour 
for our finite inhomogeneous (non-periodic !) IPL setup and what the topological markers would be.
Another direction would be to extend the current finite lattices to, principally, infinite ones 
and let the phase, according to some 'rule', evolve indefinitely. Both periodic and non-periodic
cases are then possible. These are only two out of several possible future routes of investigation 
of IPL.

\section{Acknowledgments}
\label{ACK}

The author acknowledges many inspiring discussions with F.K. Diakonos.
This work has been supported by the Cluster of Excellence
'Advanced Imaging of Matter' of the Deutsche Forschungsgemeinschaft (DFG) - EXC
2056 - project ID 390715994.

\end{document}